\documentclass[aps,twocolumn,showpacs,prb,amsmath,amssymb,nofootnoteinbib,superscriptaddress]{revtex4-2} 
\pdfoutput=1
\usepackage{graphicx,hyperref}
\usepackage{color}
\usepackage{hhline}
\usepackage{mathrsfs}
\usepackage{mathtools}
\usepackage{dcolumn}
\usepackage{bm}
\usepackage{multirow}
\usepackage{booktabs}
\usepackage{afterpage}
\usepackage{amsmath}
\usepackage{braket}
\usepackage{txfonts} 

\usepackage{bibentry}

\allowdisplaybreaks

\hypersetup{
 setpagesize=false,
 bookmarksnumbered=true,%
 bookmarksopen=true,%
 colorlinks=true,%
 linktocpage=true,
 linkcolor=blue,
 citecolor=blue,
 urlcolor=blue
}

\newcommand{\ssss}{\sigma\sigma\sigma\sigma}

\newcommand{\ssbb}{\sigma\sigma\bar{\sigma}\bar{\sigma}}

\newcommand{\ph}{\text{ph}}
\newcommand{\phb}{\overline{\text{ph}}}
\newcommand{\pp}{\text{pp}}

\begin{document}

\title{Study on the validity of IPT+parquet  method as an impurity solver in DMFT focusing on orbital fluctuations }

\author{Aira Yamada}
\affiliation{Department of Physics, The University of Osaka, 1-1 Machikaneyama, Toyonaka, Osaka 560-0043, Japan}
\author{Ryota Mizuno}\email{mizuno@presto.phys.sci.osaka-u.ac.jp}
\affiliation{Forefront Research Center, The University of Osaka, 1-1 Machikaneyama, Toyonaka, Osaka 560-0043, Japan}
\author{Masayuki Ochi}
\affiliation{Department of Physics, The University of Osaka, 1-1 Machikaneyama, Toyonaka, Osaka 560-0043, Japan}
\affiliation{Forefront Research Center, The University of Osaka, 1-1 Machikaneyama, Toyonaka, Osaka 560-0043, Japan}
\author{Kazuhiko Kuroki}
\affiliation{Department of Physics, The University of Osaka, 1-1 Machikaneyama, Toyonaka, Osaka 560-0043, Japan}
\author{Takuma Ohashi}
\affiliation{Department of Physics, The University of Osaka, 1-1 Machikaneyama, Toyonaka, Osaka 560-0043, Japan}

\date{\today}
\begin{abstract}
A breakdown of calculations with exact impurity solvers in the dynamical mean field theory in multiband systems easily occurs due to the expensive numerical cost.
To overcome this practical difficulty, three of the present authors developed an inexpensive and reliable impurity solver by combining the iterative perturbation theory (IPT) and parquet equation, and named it IPT+parquet~[R. Mizuno, \textit{et al}., Phys. Rev. B 104, 035160 (2021).].
In this study, we validate IPT+parquet focusing on the orbital fluctuation by comparing the numerically exact impurity solvers. 
We confirm that IPT+parquet can capture competition between orbital fluctuation channels, which the conventional IPT cannot capture.
\end{abstract}

\maketitle

\section{Introduction}
Strong correlation effects are crucial for various exciting phenomena such as high-temperature superconductivity and the metal-to-insulator transition.
However, it is difficult to handle the strong correlation effects correctly since they often emerge in non-perturbative regimes where both the expansions from itinerant and localized pictures break down.
In addition, there are some phenomena where the multiband nature can also be crucial: the orbital selective Mott transition~\cite{PhysRevLett.92.216402,PhysRevLett.102.126401,PhysRevLett.103.097001} and multiband superconductivity such as iron based superconductors~\cite{Kamihara2008} and multi-layer nickelate superconductors~\cite{Sun2023,PhysRevB.109.144511,Zhu2024}, and so on.
To address these complex phenomena, it is essential to correctly treat strong correlation effects in multiband systems.

The dynamical mean field theory (DMFT)~\cite{RevModPhys.68.13} is one of the most powerful methods for analyzing the strongly correlated systems.
DMFT can capture the temporal fluctuation of the effective field correctly while ignoring the spatial fluctuation.
In DMFT, we solve the lattice problem by mapping it onto the Anderson impurity problem.
Although several impurity solvers and extensions for incorporating the spatial fluctuation ignored in the single-site DMFT have been proposed so far, there has been a trade-off problem concerning the numerical costs and adaptability. 
To overcome these difficulties, three of the present authors developed inexpensive and reliable approaches for these purposes~\cite{PhysRevB.104.035160,doi:10.7566/JPSJ.91.034002,PhysRevB.111.205136}.
For the impurity solver~\cite{PhysRevB.104.035160}, the authors combined the iterative perturbation theory (IPT) and the parquet equation.
This method is called IPT+parquet.

In Ref.~\cite{PhysRevB.104.035160}, the authors applied IPT+parquet method to a single-band model, a degenerate two-band model, and a non-degenerate two-band model, and validated it by comparing the results with the numerically exact continuous-time quantum Monte Carlo method (CT-QMC).
 They demonstrated that their method achieves good agreement with CT-QMC in all cases at very low computational cost.
In particular, for the non-degenerate model—where the conventional IPT fails to produce reliable results—the IPT+parquet method shows a significant improvement over conventional IPT.
However, the benchmarks in Ref.~\cite{PhysRevB.104.035160} are limited to systems in which orbital fluctuations play only a minor role.
In multi-band systems, the importance of orbital fluctuations has been pointed out in previous studies~\cite{PhysRevB.66.165107,PhysRevLett.92.216402,PhysRevLett.102.126401,PhysRevB.83.205112}.
It is well known that conventional IPT fails to capture the effect of orbital fluctuations arising from the competition between the intra-orbital interaction $U$ and the inter-orbital interaction $U'$, namely, competing orbital fluctuation channels associated with atomic states in the $U \gg U'$ and $U \ll U'$ limits.
This is because the $U$- and $U'$-dependent terms in the second-order perturbative self-energy differ only by an overall constant factor.

Given this, in this study, we validate IPT+parquet method in systems where the orbital fluctuations play an essential role. 
We compare the IPT+parquet method with the conventinal IPT, CT-QMC and the exact diagonalization method (ED), and confirm that IPT+parquet  can correctly describe the orbital fluctuation effects, whereas the conventional IPT fails to do so.
We also show that the parquet contribution in the IPT+parquet method plays a crucial role in capturing the orbital fluctuation effects, while the orbital dependent pseudo chemical potential is essential for describing the orbital-dependent Mottness.


This paper is organized as follows. 
In Sec.~\ref{sec:Model_and_Method}, we introduce the model and methods used in this study.
We show the results in Sec.~\ref{sec:result}.
The conclusion is given in Sec.~\ref{sec:conclusion}.

\section{Model and Method}\label{sec:Model_and_Method}
\subsection{definitions}
We consider the multiband Hubbard model described by the following Hamiltonian
\begin{align}
  H 
  =&
\sum_{ij}\sum_{\alpha\beta}t_{ij,\alpha\beta}c^{\dagger}_{i\alpha}c_{j\beta} 
  +
  \dfrac{1}{4} \sum_{i} \sum_{\alpha\beta\gamma\lambda} U_{\alpha\beta\gamma\lambda} c^{\dagger}_{i\alpha}c^{\dagger}_{i\lambda}c_{i\gamma}c_{i\beta}, 
  \label{eq:2025-05-23-12-48}
\end{align}
where the subscripts with Roman letters~($i,j, ...$) indicate unit cells 
and Greek letters~($\alpha, \beta, ...$) the set of the degrees of freedom of spin, orbital, and site.
$t_{ij,\alpha\beta}$ is the hopping integral 
and 
$U_{\alpha\beta\gamma\lambda}$ is the Coulomb repulsion.
$c_{i\alpha}^{(\dagger)}$ is the annihilation (creation) operator.


The one-particle Green's function in the momentum space can be written as 
\begin{align}
  G_{\alpha\beta}(\bm{k},\tau)
  =&
  -\bigl<T c_{\bm{k}\alpha}(\tau)c^{\dagger}_{\bm{k}\beta} \bigr>
\end{align}
where
$c^{(\dagger)}(\tau)=e^{\tau H}c^{(\dagger)}e^{-\tau H}$ is the Heisenberg representation of creation (annihilation) operators.
$\braket{A}={\rm Tr}(e^{-\beta H}A)/Z$ is the statistical average of $A$
and 
$Z={\rm Tr}(e^{-\beta H})$ is the partition function.
$\bm{k}$ denotes the momentum.
The Fourier transformation in terms of the imaginary time is expressed as 
\begin{gather}
  G_{\alpha\beta}(\bm{k},\tau) = \dfrac{1}{\beta}\sum_{n} G_{\alpha\beta}(\bm{k},i\omega_{n}) e^{-i\omega_{n}\tau}
  \label{eq:2025-05-24-14-43} \\
  G_{\alpha\beta}(\bm{k},\omega_{n}) = \int_{0}^{\beta} d\tau \ G_{\alpha\beta}(\bm{k},\tau)e^{i\omega_{n}\tau}
  \label{eq:2025-05-24-14-44}
\end{gather}
where 
$\omega_{n}=(2n+1)\pi T$ with $n \in {\mathbb Z}$ is a fermionic Matsubara frequency
[$\nu_{m}=2m \pi T$ introduced later is a bosonic Matsubara frequency]. 
$G(\bm{k},i\omega_{n})$ can be derived in the following form
\begin{align}
  G(k) 
  =&
  \bigl[ (i\omega_{n} + \mu)I - \epsilon_{\bm{k}} - \Sigma(k) \bigr]^{-1},
  \label{eq:2025-05-24-14-45}
\end{align}
where 
$\mu$ is the chemical potential and 
$k=(\bm{k},i\omega_{n})$ is the generalized fermionic momentum
[$q=(\bm{q},i\nu_{m})$ introduced later denotes the generalized bosonic momentum].
$\epsilon_{\bm{k}}=N_{\bm{k}}^{-2}\sum_{ij}t_{ij}e^{i(\bm{R}_{i}-\bm{R}_{j})\cdot\bm{k}}$ is the band dispersion
and 
$\Sigma(k)$ is the self-energy.
These quantities are matrices in terms of the band index
and $I$ is the unit matrix.

The two-particle Green's function in the momentum space can be written as 
\begin{align}
  G^{(2)}_{\alpha\beta\gamma\lambda}&(\bm{k},\bm{k}',\bm{q}, \tau_{1},\tau_{2},\tau_{3}) 
  \nonumber \\
  =&
  \Bigl< T c_{\bm{k}\alpha}(\tau_{1})c^{\dagger}_{\bm{k}+\bm{q}\beta}(\tau_{2})c_{\bm{k}'+\bm{q}\lambda}(\tau_{3})c^{\dagger}_{\bm{k}'\gamma} \Bigr> .
  \label{eq:2020-10-08-23-02}
\end{align}
Fourier transformation is given by 
\begin{align}
  G^{(2)}&(\bm{k},\bm{k}',\bm{q}, \tau_{1},\tau_{2},\tau_{3}) \nonumber \\
  =&
  \dfrac{1}{\beta^{3}} \sum_{nn'm} 
  G^{(2)}(\bm{k},\bm{k}',\bm{q}, i\omega_{n},i\omega_{n'},i\nu_{m})
  \nonumber \\
  &\hspace{30pt} \times
  e^{-i\omega_{n} \tau_{1}} e^{i(\omega_{n}+\nu_{m})\tau_{2}} e^{-i(\omega_{n'}+\nu_{m})\tau_{3}} .
  \label{eq:2020-10-08-23-03}
\end{align}
Generally, we can write the two-particle Green's function as
\begin{align}
  &G^{(2)}_{\alpha\beta\gamma\lambda}(k,k',q) \nonumber \\
  &=
  G_{\alpha\beta}(k) G_{\lambda\gamma}(k') \delta_{q,0} 
  -
  G_{\alpha\gamma}(k) G_{\lambda\beta}(k+q) \delta_{kk'}
  \nonumber \\
  &+
  \hspace{-5pt}\sum_{\alpha'\beta'\gamma'\lambda'}\hspace{-5pt}
  G_{\alpha\gamma'}(k) G_{\lambda'\beta}(k+q) F_{\gamma'\lambda'\alpha'\beta'}(k,k',q) G_{\alpha'\gamma}(k') G_{\lambda\beta'}(k'+q),
  \label{eq:2020-10-08-23-04}
\end{align}
where  
$F$ is called the full vertex.
The term containing $F$ is called connected, and the other two are called disconnected.
Also, we can write the self-energy by using the full vertex as 
\begin{align}
  \Sigma_{\alpha\beta}(k) 
  =&
  \Sigma_{\alpha\beta}^{\rm HF} 
  + 
  \dfrac{1}{2}\sum_{\gamma\lambda}\sum_{k',q} [F(k,k',q) \chi_{0}(k',q) U]_{\alpha\gamma\beta\lambda} G_{\gamma\lambda}(k+q) ,
  \label{eq:2020-10-08-23-07}
\end{align}
where 
$\Sigma^{\rm HF}$ is the Hartree-Fock term and $\chi_{0}$ is the irreducible susceptibility defined below.

\subsection{Parquet equation}
\begin{figure*}[t]
  \centering
  {\includegraphics[width=180mm,clip]{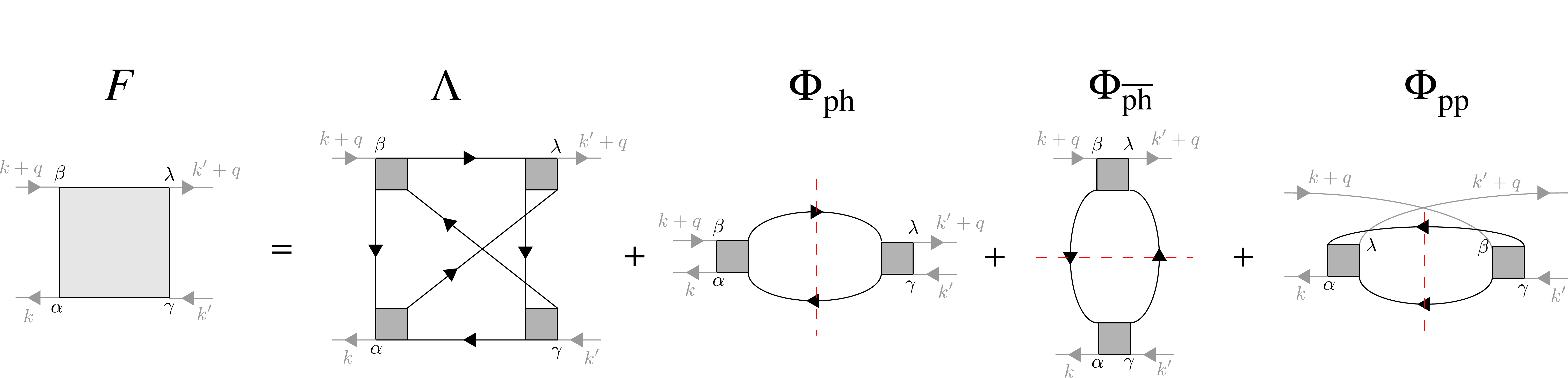}} 
  \caption{The decomposition of the full vertex.
    The full vertex can be divided into four parts:
    the fully irreducible part ($\Lambda$) and the reducible parts ($\Phi_{l}$,  $l=$ ph, ${\rm \overline{ph}}$, pp).
    The reducible vertices $\Phi_{l}$ can be divided into two parts by cutting a pair of Green's functions as indicated by the red dashed lines.
  }
  \label{fig:2020-05-09-22-21}
\end{figure*}

When we consider the diagrammatic structure of the full-vertex $F$, we define the irreducible susceptibilities concerning the following three channels (${\rm ph, \overline{ph}, pp }$):
\begin{align}
  \chi_{0,\alpha\beta\gamma\lambda}(k,k',q) =&
  \begin{cases}
    - G_{\alpha\gamma}(k)G_{\lambda\beta}(k+q) \delta_{kk'} \hspace{5pt} &(\text{ph}) \\
    G_{\alpha\beta}(k) G_{\lambda\gamma}(k') \delta_{q0}    &({\overline{\rm ph}}) \\
    G_{\alpha\gamma}(k)G_{\beta\lambda}(-k-q)\delta_{kk'}     &(\text{pp})
  \end{cases}.
  \label{eq:2024-05-09-14-46}
\end{align}
The full vertex $F$ can be divided into four parts
\begin{align}
  F(D) &= \Lambda(D) + \Phi_{\rm ph}(D) + \Phi_{\rm \overline{ph}}(D) + \Phi_{\rm pp}(P) \nonumber \\
  &= \Lambda(D) + \Phi_{\rm ph}(D) - \Phi_{\rm ph}(C) + \Phi_{\rm pp}(P),
  \label{eq:2020-05-09-22-20}
\end{align}
where
$\Phi_{l}$ $(l={\rm ph,\overline{ph},pp})$
is the set of reducible diagrams in channel $l$,
and 
$\Lambda$ is the set of fully irreducible diagrams.
Also, we introduce the notation for the set of orbital and frequency variables here as
\begin{align}
  D =& (\alpha,\beta,\gamma,\lambda), (k,k',q), \label{eq:2020-05-10-14-49} \\
  T =& (\alpha,\beta,\lambda,\gamma), (k,-q-k',q), \label{eq:2021-08-12-14-54} \\
  C =& (\alpha,\gamma,\beta,\lambda), (k,k+q,k'-k), \label{eq:2020-05-10-14-51} \\
  P =& (\alpha,\lambda,\gamma,\beta), (k,k',-q-k-k'), \label{eq:2020-05-10-14-52} \\
  X =& (\alpha,\gamma,\lambda,\beta), (k,-k'-q,k'-k). \label{eq:2020-05-10-14-53}
\end{align}
The diagrammatic representation is shown in Fig.~\ref{fig:2020-05-09-22-21}.
Since no diagram satisfies reducibility in two or more channels simultaneously, we can write
\begin{align}
  F =& \Gamma_{l} + \Phi_{l},
  \label{eq:2020-05-09-22-53} \\
  \Gamma_{l} =& \Lambda + \Phi_{l_{1}} + \Phi_{l_{2}} \hspace{10pt} (l\neq l_{1} \neq l_{2}),
  \label{eq:2020-05-09-22-54} \\
  \Phi_{l} =&
  -\Gamma_{l}\chi_{0} F = -\Gamma_{l}\chi_{l} \Gamma_{l},
\end{align}
where
$\Gamma_{l}$ is the set of diagrams irreducible in channel $l$
and 
is called the irreducible vertex in $l$.
The generalized susceptibility in channel $l$ is given by
\begin{align}
  \chi_{l} =& \chi_{0} - \chi_{0}\Gamma_{l}\chi_{l} = \chi_{0} - \chi_{0} F\chi_{0} . \label{eq:2020-05-12-14-40}
\end{align}
We omit the orbital and frequency indices in Eqs.~(\ref{eq:2020-05-09-22-53})-(\ref{eq:2020-05-12-14-40}) since it depends on the channels.
Hereafter, we may omit the index when we do not specify the channel.
From
Eqs.~(\ref{eq:2020-05-09-22-53}) to (\ref{eq:2020-05-12-14-40}), which are called the parquet equations~\cite{e_023_03_0489,PhysRevB.86.125114,Janis_1998,PhysRevB.60.11345},
we can calculate $F$ exactly if we know the exact $\Lambda$.
However, it is almost impossible to obtain the exact $\Lambda$
and the procedure to obtain $\Phi_{l}$ is computationally very expensive.
Thus, some approximations or simplifications have been proposed~\cite{doi:10.1143/JPSJ.79.094707,PhysRevB.75.165108,PhysRevB.83.035114, PhysRevB.104.035160}.

\subsection{Iterative Perturbation Theory (IPT)}
IPT~\cite{10.1143/PTP.53.970,PhysRevB.45.6479,PhysRevLett.77.131,PhysRevB.55.16132,PhysRevB.86.085133} is one of the low cost impurity solver in DMFT. 
In IPT, the correlation part of the self-energy (the self-energy subtracted by the static Hartree-Fock terms) $\Sigma^{\text{cr}}$ is approximated as 
\begin{align}
&\hspace{15pt}
\Sigma^{\text{cr}}(\omega_{n})
=
[I-B\Sigma^{\text{2nd}}(\omega_{n})]^{-1}A\Sigma^{\text{2nd}}(\omega_{n}),
\label{eq:2025-05-24-15-38} \\
&
\Sigma^{\text{2nd}}(\omega_{n})
=
T\sum_{\gamma\lambda}\sum_{m}[U\chi_{0}(\nu_{m})U]_{\alpha\gamma\beta\lambda} G_{0,\gamma\lambda}(\omega_{n}+\nu_{m}),
\label{eq:2025-05-24-15-39}\\
&\hspace{10pt}
\chi_{0,\alpha\beta\gamma\lambda}(\nu_{m})
=
-T\sum_{n}G_{0,\alpha\gamma}(\omega_{n})G_{0,\lambda\beta}(\omega_{n}+\nu_{m}),
\label{eq:2025-05-24-15-40} \\
&\hspace{15pt}
G_{0}(\omega_{n})
=
[(i\omega_{n}+\mu_{0})I - \Delta(i\omega_{n}) -\Sigma^{\text{HF}}]^{-1},
\label{eq:2025-05-24-15-41}
\end{align}
where, $\mu_{0}$, $\Delta(\omega_{n})$, and $\Sigma^{\text{HF}}$ are the pseudo chemical potential, the hybridization function, and the Hartree-Fock term in the self-energy, respectively.
We determine the parameters $A$ and $B$ such that one can reproduce the exact solution in the high-frequency and atomic limits as 
\begin{align}
    &\hspace{20pt}
    A = \dfrac{n(1-n)}{n_{0}(1-n_{0})},
    \label{eq:2025-05-25-04-15} \\
     &
     B = \dfrac{(1-2n)U + \mu_{0}-\mu}{Un(1-n)U},
    \label{eq:2025-05-25-04-16} 
\end{align}
where $n_{0}$ and $n$ are the total electron numbers obtained from the $G_{0}(\omega_{n})$ and $G(\omega_{n})$, respectively.
The expressions of $A$ and $B$ in Eqs.~(\ref{eq:2025-05-25-04-15}) and (\ref{eq:2025-05-25-04-16}) are only for the single band cases.
Several multiband extensions have been proposed~\cite{PhysRevLett.91.156402,Dasari2016}.
There are some ways to fix the pseudo chemical potential $\mu_{0}$~\cite{PhysRevLett.77.131,PhysRevB.55.16132,PhysRevB.86.085133}. 

\subsection{IPT+parquet}
Three of the present authors have extended IPT to capture the strong correlation effects correctly in the multiband systems while keeping the numerical cost low by combining IPT with the parquet equation~\cite{PhysRevB.104.035160}.
They named the method "IPT+parquet". 
In the IPT+parquet method, the correlation part of the self-energy is approximated as 
\begin{align}
    &\hspace{15pt}
    \Sigma^{\text{cr}}(\omega_{n}) 
    =
    [I-B\Sigma^{\text{cr0}}(\omega_{n})]^{-1}A\Sigma^{\text{cr0}}(\omega_{n}),
    \label{eq:2025-05-25-02-17} \\
    &
    \Sigma^{\text{cr0}}_{\alpha\beta}(\omega_{n})
    =
    T\sum_{\gamma\lambda}\sum_{n',m}[F_{0}(\omega_{n},\omega_{n'},\nu_{m})\chi_{0}(\omega_{n'},\nu_{m})U]_{\alpha\gamma\beta\lambda} \nonumber \\
    & \hspace{100pt} \times G_{0,\gamma\lambda}(\omega_{n}+\nu_{m}),
    \label{eq:2025-05-25-02-25}\\
    &\hspace{15pt}
    \chi_{0,\alpha\beta\gamma\lambda}(\omega_{n},\nu_{m})
    =
    -G_{0,\alpha\gamma}(\omega_{n})G_{0,\lambda\beta}(\omega_{n}+\nu_{m}),
    \label{eq:2025-05-25-02-27} \\
    &\hspace{20pt}
    G_{0}(\omega_{n})
    =
    [i\omega_{n}I+\mu_{0} - \Delta(i\omega_{n}) -\Sigma^{\text{HF}}]^{-1},
    \label{eq:2025-05-25-02-28}
    \\
    &
    F_{0}(\omega_{n},\omega_{n'},\nu_{m}) 
    =
    U + \Phi_{\text{ph}}(\nu_{m}) + \Phi_{\overline{\text{ph}}}(\omega_{n}-\omega_{n'}) \nonumber \\
    & \hspace{100pt} + \Phi_{\text{pp}}(\omega_{n}+\omega_{n'}+\nu_{m}),
    \label{eq:2025-05-25-02-35}
\end{align}
where $\Phi_{l}, (l=\text{ph}, \overline{\text{ph}},\text{pp})$ is the approximated reducible vertex in channel $l$, which conveys only the bosonic frequency. 
The simplified parquet method~\cite{doi:10.1143/JPSJ.79.094707,PhysRevB.104.035160} is employed to obtain the quantity $F_{0}$. 
In the IPT+parquet method, the pseudo chemical potential acquires orbital dependence and becomes $\mu_{0\alpha}$, which is fixed by the condition $n_{\alpha}=n_{0\alpha}$. Here, $n_{\alpha} = G_{\alpha\alpha}(\tau=-0)$ and $n_{0\alpha} = G_{0\alpha\alpha}(\tau=-0)$.

\section{Result}\label{sec:result}
In this section, we show the results of the IPT+parquet method.
We use the the quasiparticle weight $Z$ as probe of the correlation effectes.
The quasiparticle weight $Z$ is defind as
\begin{align}
 &\hspace{15pt}
 Z_\alpha=\left(1-\frac{\operatorname{Im}\Sigma_{\alpha\alpha}(\omega_n)}{\omega_n}\Biggr|_{\omega_n\rightarrow0}\right)^{-1}.
 \label{eq:2025-07-28-18-07}
\end{align}
$Z_{\alpha}$ is roughly proportional to the inverse of the effective mass, and $Z_\alpha=0$ corresponds to the insulating state. In this study, however, we adopt the following definition instead of Eq. (\ref{eq:2025-07-28-18-07}) for computational simplicity: 
\begin{align}
 &\hspace{15pt}
 Z_\alpha=\left(1-\frac{\operatorname{Im}\Sigma_{\alpha\alpha}(\omega_n)}{\omega_n}\Biggr|_{n=0}\right)^{-1}.
\end{align}
Also, in this study, we adopt the definition of the band-filling $n_\alpha$ [$=G_{\alpha\alpha}(\tau=-0)$] as the number of electrons per site per spin.

\begin{figure}[]
 \centering
 \includegraphics[width=240pt]{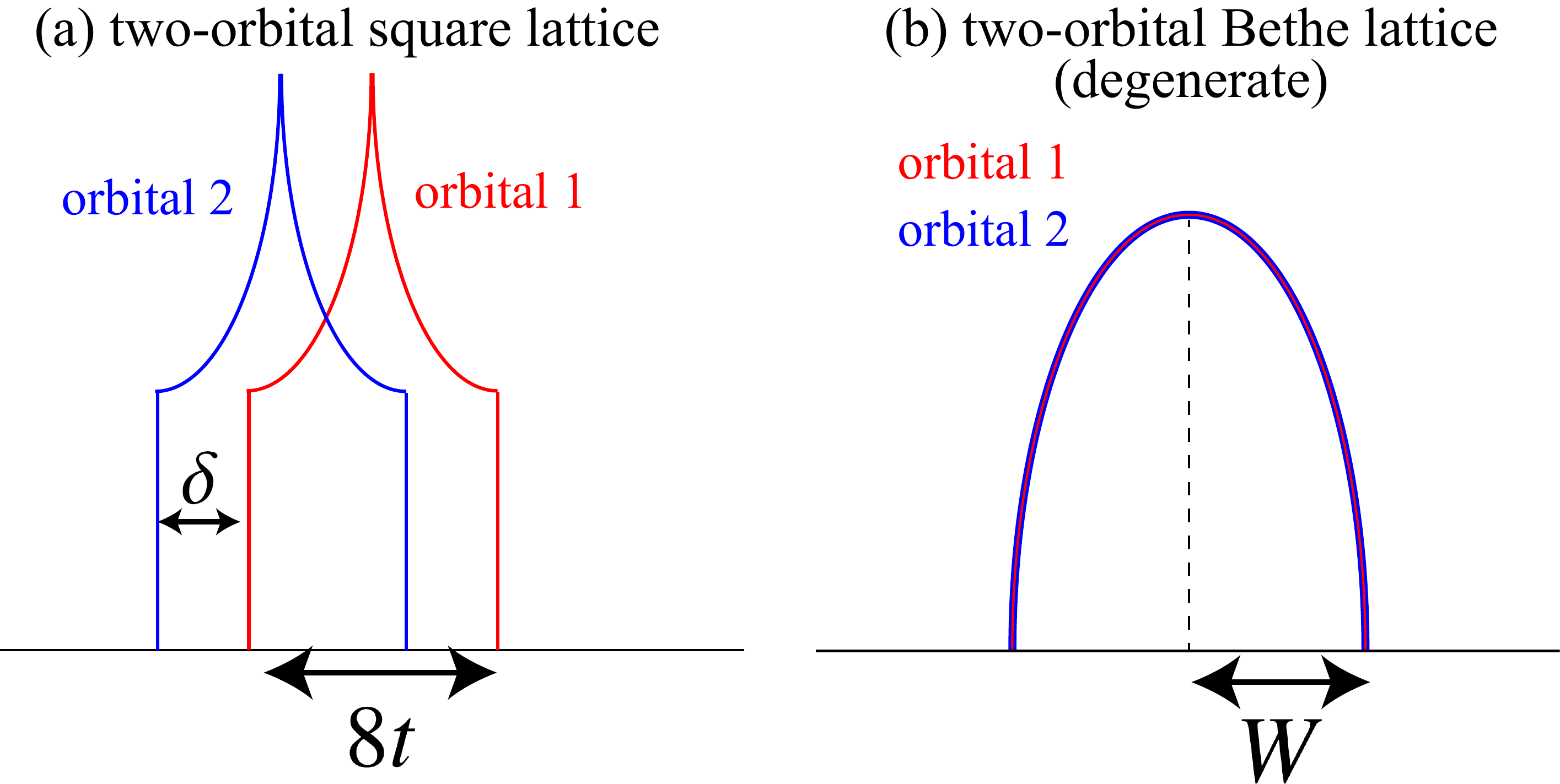}
 \caption{
 The noninteracting density of states of the models.
 Left and right panels show the density of states of two-orbital square lattice and two-orbital Bethe lattice, respectively.
 }
 \label{fig:2025-07-02-16-05}
\end{figure}

We study the two-orbital model, 
whose one-body part of the Hamiltonian is expressed as
\begin{align}
 &\hspace{15pt}
 H_0=
 \sum_{ij}\sum_{\alpha\beta}t_{ij,\alpha\beta}c^\dagger_{i\alpha}c_{j\beta}
 -\mu\sum_{i}\sum_\alpha n_{i\alpha}.
\end{align}
 The interaction part of the Hamiltonian is expressed as
\begin{align}
 &\hspace{15pt}
 H_{\rm int}=\sum_{l} Un_{l\uparrow}n_{l\downarrow}
 +\sum_{l_1< l_2}\sum_{\sigma_1\sigma_2}U'n_{l_1\sigma_1}n_{l_2\sigma_2}
 \nonumber 
 \\ & \hspace{50pt} 
+\sum_{l_1<l_2}J\bm{S}_{l_1} \cdot \bm{S}_{l_2}
+\sum_{l_1,l_2}J'
c^\dagger_{l_1\uparrow}c^\dagger_{l_1\downarrow}c_{l_2\downarrow}c_{l_2\uparrow},
\end{align}
where $l$ and $\sigma$ denote the orbital and spin, respectively.
$U^{(')}$ is the intraorbital (interorbital) interaction, and $J$ and $J'$ represent the Hund's coupling and pair hopping, respectively. 
The interaction matrices in charge and spin channels $U^{c}$ and $U^{s}$ are defined as 
\begin{align}
    U^{c}_{l_{1}l_{2}l_{3}l_{4}} 
    =& 
    U^{\ssss}_{l_{1}l_{2}l_{3}l_{4}}  + U^{\ssbb}_{l_{1}l_{2}l_{3}l_{4}}  \label{eq:2026-01-28-14-56} \\
    U^{s}_{l_{1}l_{2}l_{3}l_{4}}  
    =& 
    U^{\ssss}_{l_{1}l_{2}l_{3}l_{4}}  - U^{\ssbb}_{l_{1}l_{2}l_{3}l_{4}}  \label{eq:2026-01-28-14-57}
\end{align}
where $U^{\sigma_{1}\sigma_{2}\sigma_{3}\sigma_{4}}_{l_{1}l_{2}l_{3}l_{4}} = U_{\alpha\beta\gamma\lambda}$ denotes a representation in which the composite indices $(\alpha,\beta,\gamma,\lambda)$ are divided into spin $\sigma_{i}$ and orbital $l_{i}$ indices. 
The symbol $\bar{\sigma}$ represents the spin opposite to $\sigma$.
Then the interaction matrices in the charge and spin channels in this model are expressed as
\begin{align}
(U^c_{l_1 l_2 l_3 l_4},U^s_{l_1 l_2 l_3 l_4})
=&
  \begin{cases}
  (U,U)&(l_1=l_2=l_3=l_4) \\
  (2U'-J,J)&(l_1=l_2\neq l_3=l_4) \\
  (2J-U',U')&(l_1=l_3 \neq l_2=l_4) \\
  (J',J')&(l_1=l_4 \neq l_2=l_3) \\
  \end{cases}
\end{align}
We performed calculations using two models: a two-orbital two-dimensional square lattice with equal bandwidths but different orbital levels, and a two-orbital Bethe lattice with equal bandwidths and orbital levels.
Fig. \ref{fig:2025-07-02-16-05} shows the noninterecting density of states of the models which we study.
Here, $t$ denotes the nearest-neighbor hopping amplitude,
$\delta$ is the on-site energy difference, and 
$W$ represents the half bandwidth of the density of states at
$U=0$.

\subsection{Two-orbital square lattice}
\begin{figure}[]
 \centering
 \includegraphics[width=240pt]{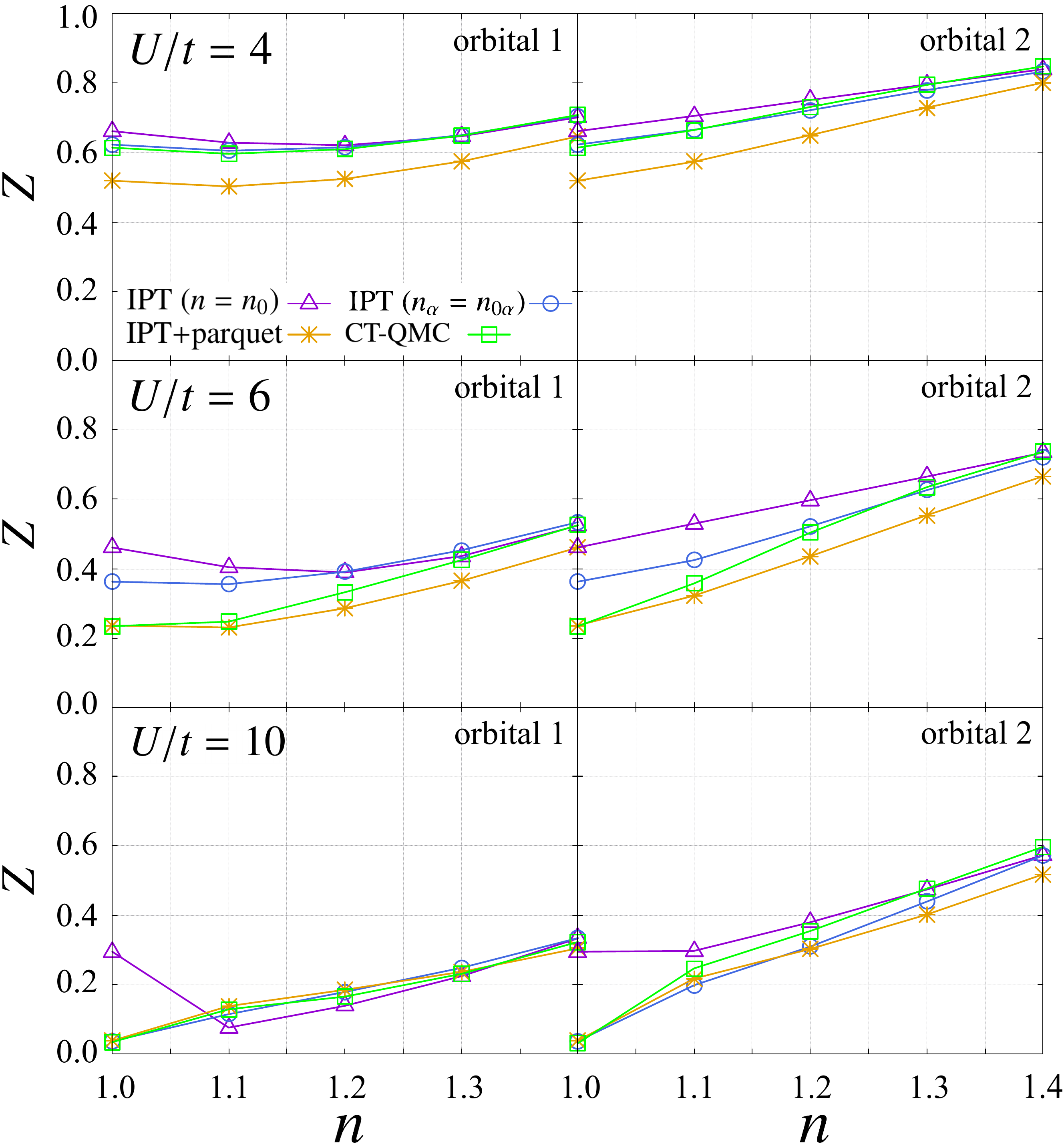}
 \caption{
 The band filling $n$ dependence of the quasiparticle weight $Z$ of the two-orbital square lattice model.
 The temperature is $T/t=0.2$ and the on-site energy difference is $\delta/t=1.6$.
 Triangle, circle, asterisk, and square symbols (purple, blue, khaki, and green lines) represent ${\rm IPT}_{n=n_0}$, ${\rm IPT}_{n_\alpha=n_{0\alpha}}$, IPT+parquet, and CT-QMC, respectively.
 Top, middle, and bottom panels show the quasiparticle weight $Z$ at $U/t=4,6,10$, respectively.
 The left and right panels indicate orbital 1 and orbital 2, respectively.
 }
 \label{fig:2025-06-20-16-34}
\end{figure}

We consider the two-orbital square lattice model which has only the intraorbital nearest-neighbor hopping.
Here, we compare three impurity solvers: IPT+parquet, IPT, and CT-QMC. 
In the IPT calculation, we use two different conditions for determining the chemical potential.
In one calculation, the condition $n = n_{0}$ is used (${\rm IPT}_{n=n_0}$), while in the other calculation, the condition $n_{\alpha} = n_{0\alpha}$ is imposed (${\rm IPT}_{n_{\alpha}=n_{0\alpha}}$).
In the CT-QMC calculation, we use the \texttt{CTHYB} package~\cite{PhysRevLett.97.076405, PhysRevB.74.155107, SETH2016274, triqs_ctqmc_gull, lewin_thesis, PhysRevB.84.075145} based on the \texttt{TRIQS} library ~\cite{PARCOLLET2015398}. 
We set $t_1=t_2=t$, where $t_\alpha \ (\alpha=1,2)$ is the nearest-neighbor hopping of orbital $\alpha$ and $t$ is the unit of energy. 
The interactions are given  by $U'=U-2J$ and $J=J'=U/4$. We take a 32 $\times$ 32 $k$-mesh and 4096 Matsubara frequencies. 
We fix the temperature $T/t=0.2$, and the on-site energy difference $\delta/t=1.6$.
Fig.~\ref{fig:2025-06-20-16-34} shows the quasiparticle weight $Z$ obtained by four methods: IPT+parquet, ${\rm IPT}_{n=n_0}$, ${\rm IPT}_{n_\alpha=n_{0\alpha}}$ and CT-QMC as a function of the filling $n$, where half-filling corresponds to $n=1$, for several interaction strengths $U$. 
Overall, ${\rm IPT}_{n=n_0}$ shows good agreement with CT-QMC in the weakly correlated regime (small $U$, away from half-filling), whereas IPT+parquet performs better in the strongly correlated regime (large $U$, near half-filling).
${\rm IPT}_{n_\alpha=n_{0\alpha}}$ provides intermediate results between ${\rm IPT}_{n=n_0}$ and IPT+parquet.
However, when $U/t=10$ in the vicinity of half-filling,  ${\rm IPT}_{n=n_0}$ deviates significantly from the CT-QMC results.
This indicates that ${\rm IPT}_{n=n_0}$ fails to correctly capture the multi-orbital Mottness.


The IPT+parquet method introduced in Ref.~\cite{PhysRevB.104.035160} incorporates both reducible vertices $\Phi_{l}$ and orbital-dependent pseudochemical potentials $\mu_{0\alpha}$~(the condition $n_{\alpha}=n_{0\alpha}$), and has been shown to achieve good agreement with CT-QMC, particularly in the strongly correlated regime.
Here, we examine ${\rm IPT}_{n_{\alpha} = n_{0\alpha}}$ as a simplified scheme in which only the orbital-dependent pseudochemical potential $\mu_{0\alpha}$ is introduced, without includeing the reducible vertices.
Remarkably, this simplified approach still reproduces the CT-QMC results with good accuracy. 
Furthermore, in the weakly correlated regime, it even outperforms IPT+parquet, suggesting that the inclusion of reducible vertices may lead to an overestimation of interaction effects in this regime.
In contrast, ${\rm IPT}_{{n} = n_{0}}$ (the conventional IPT with a single pseudochemical potential $\mu_{0}$) shows significant deviations near half-filling and fails to capture orbital-dependent correlation effects.
These results indicate that the introduction of orbital-dependent pseudochemical potentials $\mu_{0\alpha}$ and the condition $n_{\alpha}=n_{0\alpha}$ are crucial for accurately capturing the orbital dependent strong correlation effects.

\subsection{Two-orbital Bethe lattice}
\begin{figure*}[]
 \centering
 \includegraphics[width=430pt]{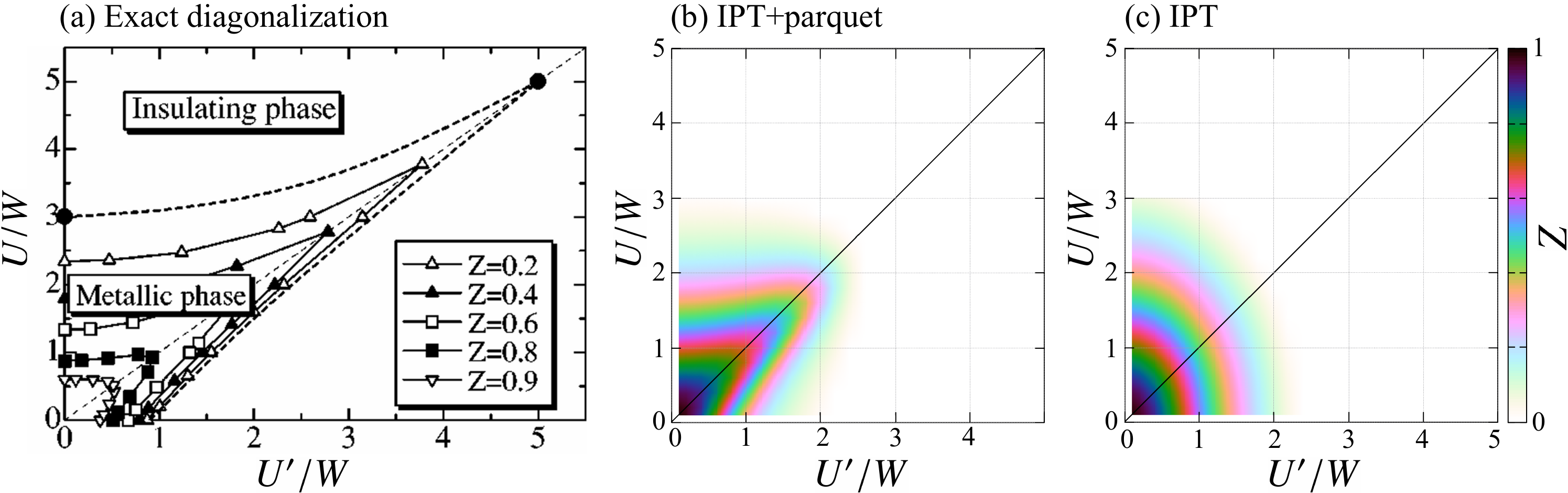}
 \caption{
 The interaction strength ($U,U'$) dependence of the quasiparticle weight $Z$ of the two-orbital Bethe lattice model obtained by (a) Exact diagonalization, (b) IPT+parquet, and (c) IPT, respectively.
 The Hund coupling and pair hopping are fixed as $J=J'=0$. 
 The temperatures are (a) $T/W=0$ and (b), (c)  $T/W=0.005$, respectively.
 Left figure is taken from Ref.~\cite{PhysRevB.66.165107}.
 }
 \label{fig:2025-06-20-16-33}
\end{figure*}

\begin{figure}[]
 \centering
 \includegraphics[width=240pt]{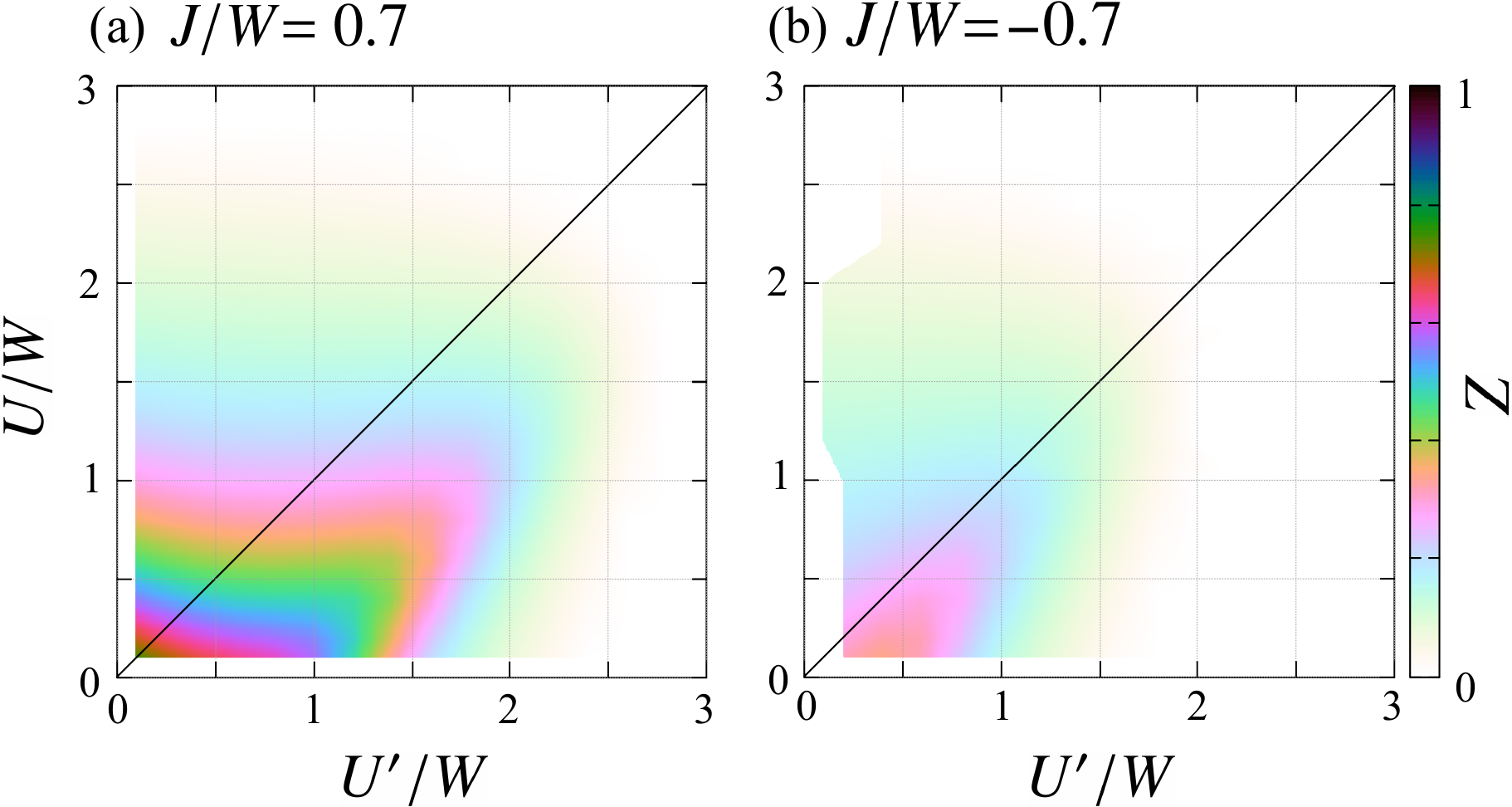}
 \caption{The interaction strength ($U,U'$) dependence of the quasiparticle weight $Z$ of the two-orbital Bethe lattice model obtained by IPT+parquet.
 The Hund coupling is (a) $J/W=0.7$ and (b) $J/W=-0.7$, respectively.
 The pair hopping is $J'=0$. 
 The temperature is $T/t=0.005$.
 }
 \label{fig:2025-07-02-15-34}
\end{figure}

We study the two-orbital Bethe lattice model in which two bands with identical bandwidth and energy levels exist.
We fixed the temperature at $T/W=0.005$ and the filling at $n=1$.
Here, $W$ denotes the half bandwidth of the density of states at $U=0$.
We take 4096 Matsubara frequencies.
Fig. \ref{fig:2025-06-20-16-33} shows the interaction strength $U$ and $U'$ dependence of the quasiparticle weight $Z$ obtained by three methods: exact diagonalization, IPT+parquet, and IPT with $J=J'=0$.
Since 
the two orbitals are equivalent, there is no distinction between the two conditions, $n=n_0$ and $n_\alpha=n_{0\alpha}$; hence, we omit the subscript indicating the IPT condition.
It is well known that the metallic state is stabilized when $U'$ approaches $U$.
This behavior can be understood as arising from orbital fluctuations, namely, the competition between atomic states in the $U \gg U'$ and $U \ll U'$ limits.
This tendency has been demonstrated by the exact diagonalization result of  Ref.~\cite{PhysRevB.66.165107}, shown in Fig. \ref{fig:2025-06-20-16-33} (a).
The conventional IPT, shown in Fig.~\ref{fig:2025-06-20-16-33}(c), cannot reproduce this tendency. 
In contrast, the IPT+parquet approach qualitatively improves upon the conventional IPT. 
As shown in Fig.~\ref{fig:2025-06-20-16-33}~(b), for a given $U$, 
the quasiparticle weight increases as $U'$ approaches $U$, indicating the stabilization of the metallic phase. 
Fig. \ref{fig:2025-07-02-15-34} shows the interaction strength $U$ and $U'$ dependence of the quasiparticle weight $Z$ obtained by IPT+parquet with $J'=0$.
Fig. \ref{fig:2025-07-02-15-34} (a) and (b) show the results with $J/W=0.7$ and $J/W=-0.7$ respectively.
It is evident from both figures that for $U>U'$ the quasiparticle weight decreases as the magnitude of the exchange interaction $|J|$ increases.
The results obtained in this study are in good agreement with the results presented in Ref.~\cite{PhysRevB.66.165107}.
For the case of $J<0$, however, the calculations did not converge properly when using exactly the same value of $J$ as Ref.~\cite{PhysRevB.66.165107}, so a slightly different value was used.

In this case, 
the difference between IPT+parquet and IPT lies in whether the reducible vertices $ \Phi_{\rm ph}$, $ \Phi_{\phb}$, and $ \Phi_{\rm pp}$ in Eq. (\ref{eq:2025-05-25-02-35}) are included or not.
While the conventional IPT fails to describe the stabilization of the metallic phase in the region where $U\sim U'$, the IPT+parquet method captures it accurately.
Hence, this result clarifies that the reducible vertices play an essential role in capturing two competing orbital fluctuation channels associated with atomic states in the $U \gg U'$ and $U \ll U'$ limits.
Even when the exchange coupling $J$ is introduced, the results obtained by IPT+parquet approach accurately reproduce the key features observed in exact diagonalization results of Ref.~\cite{PhysRevB.66.165107}.
The difficulty in achieving convergence for $J<0$ originates from limitations inherent in the simplified parquet method used to evaluate the reducible vertices.
Improving the treatment of the reducible vertices and extending the calculations to the $J<0$ region remain important subjects for future work.



%

\section{Conclusion}\label{sec:conclusion}
In the present work, we have studied the validity of IPT + parquet method focusing on the orbital fluctuation.
We compared the results obtained from CT-QMC, IPT+parquet, and two variants of IPT that differ in the conditions used to determine the pseudo chemical potential $\mu_{0}$.
As a result, we found that the reducible vertex enables IPT+parquet to capture the competing orbital fluctuation channels associated with atomic states in the $U \gg U'$ and $U \ll U'$ limits.  
Additionally, we found that when the competition between the orbital fluctuation channels is not strong, IPT-type methods can capture orbital-dependent correlation effects solely through the orbital degrees of freedom in the pseudo chemical potential \(\mu_{0\alpha}\), without involving the reducible vertex terms.  
Neither of these features can be achieved by conventional interpolation-type IPT extensions.


\begin{acknowledgements}
  This study has been supported by Grant-in-Aid for Early-Career Scientists (Grant No.JP23K13061), JST FOREST Program (Grant No. JPMJFR212P), Grant-in-Aid for Scientific Research B (Grant No. JP24K01333), and Grant-in-Aid for Transformative Research Areas (Grant No.25H01252).

\end{acknowledgements}

\appendix

\section{How to fix the irreducible vertex $\Gamma$ in the simplified parquet method}
We use a different procedure from that of Ref.~\cite{PhysRevB.104.035160} to fix the constant irreducible vertex $\Gamma_{l}$ corresponding to $\hat{\tilde{\Lambda}}$ in Ref.~\cite{PhysRevB.104.035160}, within the simplified parquet scheme employed in the IPT+parquet method. 
Using the procedure of Ref.~\cite{PhysRevB.104.035160}, the IPT+parquet method becomes unstable when the interaction significantly deviates from the condition $U'=U-2J$ with $J=U/4$. 
We therefore modify the equation for determining $\Gamma_{l}$ as follows.
\begin{align}
&\Gamma_{\ph,\alpha\beta\gamma\lambda} \nonumber \\
&=
\frac{
\sum_{k k' q}
\left( \chi_0(k', q)\,\chi_0(k, q) \right)_{\gamma\lambda\alpha\beta}
\left[
\Lambda_{\ph,\alpha\beta\gamma\lambda}
- \Phi_{\ph}(C)
+ \Phi_{\pp}(P)
\right]
}{
\sum_q
\left( \chi_0(q)\,\chi_0(q) \right)_{\gamma\lambda\alpha\beta}
}
,
\label{eq:2026-04-15-01-32} \\
&\Gamma_{\pp,\alpha\beta\gamma\lambda} \nonumber \\
&=
\frac{
\sum_{k k' q}
\left( \chi_0(k', q)\,\chi_0(k, q) \right)_{\gamma\lambda\alpha\beta}
\left[
\Lambda_{\pp,\alpha\beta\gamma\lambda}
- \Phi_{\ph}(X)
+ \Phi_{\ph}(P)
\right]
}{
\sum_q
\left( \chi_0(q)\,\chi_0(q) \right)_{\gamma\lambda\alpha\beta}
},
\label{eq:2026-04-15-01-33} 
\end{align}
where $\chi_{0}$ is the irreducible susceptibility in each channel, as defined in Eq.~(\ref{eq:2024-05-09-14-46}).
$\Lambda_{l}$ and $\Phi_{l}$ denote the fully irreducible and reducible vertex in channel $l$, respectively.
The notations $C,X$, and $P$ represent the set of orbital variables and momenta, as defined in Eqs.~(\ref{eq:2020-05-10-14-49})-(\ref{eq:2020-05-10-14-53}).
Using Eqs.~(\ref{eq:2026-04-15-01-32}) and (\ref{eq:2026-04-15-01-33}), 
each component of $\Gamma_{l,\alpha\beta\gamma\lambda}$ is determined independently, so as to satisfy the condition $\mathrm{Tr} (\chi_{0}\Gamma_{l}\chi_{0}) = \mathrm{Tr}(\chi_{0}(\Lambda_{l}+\Phi_{l_{1}} + \Phi_{l_{2}})\chi_{0}), \ (l\neq l_{1} \neq l_{2})$, where $\mathrm{Tr}A = \sum_{kk'q}\sum_{\alpha\beta}A_{\alpha\beta\alpha\beta}(k,k',q)$.
In contrast, in Ref.~\cite{PhysRevB.104.035160}, $\Gamma_{l}$ is determined by multiplying a single renormalization factor $z_{l}$ to the bare vertex. 
With this modification, IPT+parquet becomes stable even when the condition $U'=U-2J, J=U/4$ is largely violated, though IPT+parquet tends to overestimate correlation effects in the small $U$ region.

\bibliography{reference}

@article{PhysRevB.86.125114,
  title = {Local electronic correlation at the two-particle level},
  author = {Rohringer, G. and Valli, A. and Toschi, A.},
  journal = {Phys. Rev. B},
  volume = {86},
  issue = {12},
  pages = {125114},
  numpages = {26},
  year = {2012},
  month = {Sep},
  publisher = {American Physical Society},
  doi = {10.1103/PhysRevB.86.125114},
  url = {https://link.aps.org/doi/10.1103/PhysRevB.86.125114}
}

@article{e_023_03_0489,
  title={Possibility of superconductivity type phenomena in a one-dimensional system},
  author={Yu. A. Bychkov and L. P. Gor'kov and I. E. Dzyaloshinski},
  journal={Journal of Experimental and Theoretical Physics},
  year={1966},
  volume={50},
  pages={738-758},
  url="http://jetp.ras.ru/cgi-bin/e/index/r/50/3/p738?a=list",

}

@article{Janis_1998,
doi = {10.1088/0953-8984/10/13/010},
url = {https://dx.doi.org/10.1088/0953-8984/10/13/010},
year = {1998},
month = {apr},
publisher = {},
volume = {10},
number = {13},
pages = {2915},
author = {V Janis},
title = {The Hubbard model at intermediate coupling: renormalization of the interaction strength},
journal = {Journal of Physics: Condensed Matter}
}

@article{PhysRevB.60.11345,
  title = {{Stability of self-consistent solutions for the Hubbard model at intermediate and strong coupling}},
  author = {Jani\ifmmode \check{s}\else \v{s}\fi{}, V.},
  journal = {Phys. Rev. B},
  volume = {60},
  issue = {16},
  pages = {11345--11360},
  numpages = {0},
  year = {1999},
  month = {Oct},
  publisher = {American Physical Society},
  doi = {10.1103/PhysRevB.60.11345},
  url = {https://link.aps.org/doi/10.1103/PhysRevB.60.11345}
}

@article{doi:10.1143/JPSJ.79.094707,
author = {Kusunose ,Hiroaki},
title = {{Self-Consistent Fluctuation Theory for Strongly Correlated Electron Systems}},
journal = {Journal of the Physical Society of Japan},
volume = {79},
number = {9},
pages = {094707},
year = {2010},
doi = {10.1143/JPSJ.79.094707},
URL = {https://doi.org/10.1143/JPSJ.79.094707},
eprint = {https://doi.org/10.1143/JPSJ.79.094707}
}

@article{PhysRevB.75.165108,
  title = {{Analytic impurity solver with Kondo strong-coupling asymptotics}},
  author = {Jani\ifmmode \check{s}\else \v{s}\fi{}, V. and Augustinsk\'y, P.},
  journal = {Phys. Rev. B},
  volume = {75},
  issue = {16},
  pages = {165108},
  numpages = {7},
  year = {2007},
  month = {Apr},
  publisher = {American Physical Society},
  doi = {10.1103/PhysRevB.75.165108},
  url = {https://link.aps.org/doi/10.1103/PhysRevB.75.165108}
}

@article{PhysRevB.83.035114,
  title = {Multiorbital simplified parquet equations for strongly correlated electrons},
  author = {Augustinsk\'y, Pavel and Jani\ifmmode \check{s}\else \v{s}\fi{}, V\'aclav},
  journal = {Phys. Rev. B},
  volume = {83},
  issue = {3},
  pages = {035114},
  numpages = {13},
  year = {2011},
  month = {Jan},
  publisher = {American Physical Society},
  doi = {10.1103/PhysRevB.83.035114},
  url = {https://link.aps.org/doi/10.1103/PhysRevB.83.035114}
}

@article{PhysRevB.104.035160,
  title = {{Development of an efficient impurity solver in dynamical mean field theory for multiband systems: Iterative perturbation theory combined with parquet equations}},
  author = {Mizuno, Ryota and Ochi, Masayuki and Kuroki, Kazuhiko},
  journal = {Phys. Rev. B},
  volume = {104},
  issue = {3},
  pages = {035160},
  numpages = {15},
  year = {2021},
  month = {Jul},
  publisher = {American Physical Society},
  doi = {10.1103/PhysRevB.104.035160},
  url = {https://link.aps.org/doi/10.1103/PhysRevB.104.035160}
}

@article{10.1143/PTP.53.970,
    author = {Yamada, Kosaku},
    title = {{Perturbation Expansion for the Anderson Hamiltonian. II}},
    journal = {Progress of Theoretical Physics},
    volume = {53},
    number = {4},
    pages = {970-986},
    year = {1975},
    month = {04},
    issn = {0033-068X},
    doi = {10.1143/PTP.53.970},
    url = {https://doi.org/10.1143/PTP.53.970},
    eprint = {https://academic.oup.com/ptp/article-pdf/53/4/970/5428364/53-4-970.pdf},
}

@article{PhysRevB.45.6479,
  title = {Hubbard model in infinite dimensions},
  author = {Georges, Antoine and Kotliar, Gabriel},
  journal = {Phys. Rev. B},
  volume = {45},
  issue = {12},
  pages = {6479--6483},
  numpages = {0},
  year = {1992},
  month = {Mar},
  publisher = {American Physical Society},
  doi = {10.1103/PhysRevB.45.6479},
  url = {https://link.aps.org/doi/10.1103/PhysRevB.45.6479}
}

@article{PhysRevLett.77.131,
  title = {{New Iterative Perturbation Scheme for Lattice Models with Arbitrary Filling}},
  author = {Kajueter, Henrik and Kotliar, Gabriel},
  journal = {Phys. Rev. Lett.},
  volume = {77},
  issue = {1},
  pages = {131--134},
  numpages = {0},
  year = {1996},
  month = {Jul},
  publisher = {American Physical Society},
  doi = {10.1103/PhysRevLett.77.131},
  url = {https://link.aps.org/doi/10.1103/PhysRevLett.77.131}
}

@article{PhysRevB.55.16132,
  title = {{Interpolating self-energy of the infinite-dimensional Hubbard model: Modifying the iterative perturbation theory}},
  author = {Potthoff, M. and Wegner, T. and Nolting, W.},
  journal = {Phys. Rev. B},
  volume = {55},
  issue = {24},
  pages = {16132--16142},
  numpages = {0},
  year = {1997},
  month = {Jun},
  publisher = {American Physical Society},
  doi = {10.1103/PhysRevB.55.16132},
  url = {https://link.aps.org/doi/10.1103/PhysRevB.55.16132}
}

@article{PhysRevB.86.085133,
  title = {{Benchmark of a modified iterated perturbation theory approach on the fcc lattice at strong coupling}},
  author = {Arsenault, Louis-Fran{\c{c}ois} and S\'emon, Patrick and Tremblay, A.-M. S.},
  journal = {Phys. Rev. B},
  volume = {86},
  issue = {8},
  pages = {085133},
  numpages = {16},
  year = {2012},
  month = {Aug},
  publisher = {American Physical Society},
  doi = {10.1103/PhysRevB.86.085133},
  url = {https://link.aps.org/doi/10.1103/PhysRevB.86.085133}
}

@article{RevModPhys.68.13,
  title = {{Dynamical mean-field theory of strongly correlated fermion systems and the limit of infinite dimensions}},
  author = {Georges, Antoine and Kotliar, Gabriel and Krauth, Werner and Rozenberg, Marcelo J.},
  journal = {Rev. Mod. Phys.},
  volume = {68},
  issue = {1},
  pages = {13--125},
  numpages = {0},
  year = {1996},
  month = {Jan},
  publisher = {American Physical Society},
  doi = {10.1103/RevModPhys.68.13},
  url = {https://link.aps.org/doi/10.1103/RevModPhys.68.13}
}

@article{PhysRevLett.91.156402,
  title = {{Orbital Switching and the First-Order Insulator-Metal Transition in Paramagnetic $\text{V}_{2}\text{O}_{3}$}},
  author = {Laad, M. S. and Craco, L. and M\"uller-Hartmann, E.},
  journal = {Phys. Rev. Lett.},
  volume = {91},
  issue = {15},
  pages = {156402},
  numpages = {4},
  year = {2003},
  month = {Oct},
  publisher = {American Physical Society},
  doi = {10.1103/PhysRevLett.91.156402},
  url = {https://link.aps.org/doi/10.1103/PhysRevLett.91.156402}
}

@Article{Dasari2016,
author={Dasari, Nagamalleswararao
and Mondal, Wasim Raja
and Zhang, Peng
and Moreno, Juana
and Jarrell, Mark
and Vidhyadhiraja, N. S.},
title={{A multi-orbital iterated perturbation theory for model Hamiltonians and real material-specific calculations of correlated systems}},
journal={The European Physical Journal B},
year={2016},
month={Sep},
day={21},
volume={89},
number={9},
pages={202},
issn={1434-6036},
doi={10.1140/epjb/e2016-70133-4},
url={https://doi.org/10.1140/epjb/e2016-70133-4}
}

@article{PhysRevB.66.165107,
  title = {{Stability of a metallic state in the two-orbital Hubbard model}},
  author = {Koga, Akihisa and Imai, Yoshiki and Kawakami, Norio},
  journal = {Phys. Rev. B},
  volume = {66},
  issue = {16},
  pages = {165107},
  numpages = {6},
  year = {2002},
  month = {Oct},
  publisher = {American Physical Society},
  doi = {10.1103/PhysRevB.66.165107},
  url = {https://link.aps.org/doi/10.1103/PhysRevB.66.165107}
}

@article{PhysRevLett.97.076405,
  title = {{Continuous-Time Solver for Quantum Impurity Models}},
  author = {Werner, Philipp and Comanac, Armin and de' Medici, Luca and Troyer, Matthias and Millis, Andrew J.},
  journal = {Phys. Rev. Lett.},
  volume = {97},
  issue = {7},
  pages = {076405},
  numpages = {4},
  year = {2006},
  month = {Aug},
  publisher = {American Physical Society},
  doi = {10.1103/PhysRevLett.97.076405},
  url = {https://link.aps.org/doi/10.1103/PhysRevLett.97.076405}
}

@article{PhysRevB.74.155107,
  title = {{Hybridization expansion impurity solver: General formulation and application to Kondo lattice and two-orbital models}},
  author = {Werner, Philipp and Millis, Andrew J.},
  journal = {Phys. Rev. B},
  volume = {74},
  issue = {15},
  pages = {155107},
  numpages = {13},
  year = {2006},
  month = {Oct},
  publisher = {American Physical Society},
  doi = {10.1103/PhysRevB.74.155107},
  url = {https://link.aps.org/doi/10.1103/PhysRevB.74.155107}
}

@article{SETH2016274,
title = {{TRIQS/CTHYB: A continuous-time quantum Monte Carlo hybridisation expansion solver for quantum impurity problems}},
journal = {Computer Physics Communications},
volume = {200},
pages = {274-284},
year = {2016},
issn = {0010-4655},
doi = {https://doi.org/10.1016/j.cpc.2015.10.023},
url = {https://www.sciencedirect.com/science/article/pii/S001046551500404X},
author = {Priyanka Seth and Igor Krivenko and Michel Ferrero and Olivier Parcollet}
}

@article{PhysRevB.84.075145,
  title = {{Orthogonal polynomial representation of imaginary-time Green's functions}},
  author = {Boehnke, Lewin and Hafermann, Hartmut and Ferrero, Michel and Lechermann, Frank and Parcollet, Olivier},
  journal = {Phys. Rev. B},
  volume = {84},
  issue = {7},
  pages = {075145},
  numpages = {13},
  year = {2011},
  month = {Aug},
  publisher = {American Physical Society},
  doi = {10.1103/PhysRevB.84.075145},
  url = {https://link.aps.org/doi/10.1103/PhysRevB.84.075145}
}

@article{PARCOLLET2015398,
title = {{TRIQS: A toolbox for research on interacting quantum systems}},
journal = {Computer Physics Communications},
volume = {196},
pages = {398-415},
year = {2015},
issn = {0010-4655},
doi = {https://doi.org/10.1016/j.cpc.2015.04.023},
url = {https://www.sciencedirect.com/science/article/pii/S0010465515001666},
author = {Olivier Parcollet and Michel Ferrero and Thomas Ayral and Hartmut Hafermann and Igor Krivenko and Laura Messio and Priyanka Seth}
}

@article{PhysRevLett.92.216402,
  title = {{Orbital-Selective} {Mott} {Transitions} in the {Degenerate Hubbard Model}},
  author = {Koga, Akihisa and Kawakami, Norio and Rice, T. M. and Sigrist, Manfred},
  journal = {Phys. Rev. Lett.},
  volume = {92},
  issue = {21},
  pages = {216402},
  numpages = {4},
  year = {2004},
  month = {May},
  publisher = {American Physical Society},
  doi = {10.1103/PhysRevLett.92.216402},
  url = {https://link.aps.org/doi/10.1103/PhysRevLett.92.216402}
}

@article{PhysRevLett.102.126401,
  title = {{Orbital-Selective Mott Transition out of Band Degeneracy Lifting}},
  author = {de' Medici, Luca and Hassan, S. R. and Capone, Massimo and Dai, Xi},
  journal = {Phys. Rev. Lett.},
  volume = {102},
  issue = {12},
  pages = {126401},
  numpages = {4},
  year = {2009},
  month = {Mar},
  publisher = {American Physical Society},
  doi = {10.1103/PhysRevLett.102.126401},
  url = {https://link.aps.org/doi/10.1103/PhysRevLett.102.126401}
}

@article{PhysRevLett.103.097001,
  title = {{Observation of a Novel Orbital Selective Mott Transition in $\text{Ca}_{1.8}\text{Sr}_{0.2}\text{RuO}_{4}$}},
  author = {Neupane, M. and Richard, P. and Pan, Z.-H. and Xu, Y.-M. and Jin, R. and Mandrus, D. and Dai, X. and Fang, Z. and Wang, Z. and Ding, H.},
  journal = {Phys. Rev. Lett.},
  volume = {103},
  issue = {9},
  pages = {097001},
  numpages = {4},
  year = {2009},
  month = {Aug},
  publisher = {American Physical Society},
  doi = {10.1103/PhysRevLett.103.097001},
  url = {https://link.aps.org/doi/10.1103/PhysRevLett.103.097001}
}

@article{PhysRevB.83.205112,
  title = {{Hund's coupling and its key role in tuning multiorbital correlations}},
  author = {de' Medici, Luca},
  journal = {Phys. Rev. B},
  volume = {83},
  issue = {20},
  pages = {205112},
  numpages = {11},
  year = {2011},
  month = {May},
  publisher = {American Physical Society},
  doi = {10.1103/PhysRevB.83.205112},
  url = {https://link.aps.org/doi/10.1103/PhysRevB.83.205112}
}

@Article{Kamihara2008,
author={Kamihara, Yoichi
and Watanabe, Takumi
and Hirano, Masahiro
and Hosono, Hideo},
title={{Iron-Based Layered Superconductor $\text{La}[\text{O}_{1-x}\text{F}_{x}]\text{FeAs} (x=0.05-0.12)$ with $T_{c} = 26 \text{K}$}},
journal={Journal of the American Chemical Society},
year={2008},
month={Mar},
day={01},
publisher={American Chemical Society},
volume={130},
number={11},
pages={3296-3297},
issn={0002-7863},
doi={10.1021/ja800073m},
url={https://doi.org/10.1021/ja800073m}
}

@Article{Sun2023,
author={Sun, Hualei
and Huo, Mengwu
and Hu, Xunwu
and Li, Jingyuan
and Liu, Zengjia
and Han, Yifeng
and Tang, Lingyun
and Mao, Zhongquan
and Yang, Pengtao
and Wang, Bosen
and Cheng, Jinguang
and Yao, Dao-Xin
and Zhang, Guang-Ming
and Wang, Meng},
title={{Signatures of superconductivity near 80{\thinspace}{K} in a nickelate under high pressure}},
journal={Nature},
year={2023},
month={Sep},
day={01},
volume={621},
number={7979},
pages={493-498},
issn={1476-4687},
doi={10.1038/s41586-023-06408-7},
url={https://doi.org/10.1038/s41586-023-06408-7}
}

@Article{Zhu2024,
author={Zhu, Yinghao
and Peng, Di
and Zhang, Enkang
and Pan, Bingying
and Chen, Xu
and Chen, Lixing
and Ren, Huifen
and Liu, Feiyang
and Hao, Yiqing
and Li, Nana
and Xing, Zhenfang
and Lan, Fujun
and Han, Jiyuan
and Wang, Junjie
and Jia, Donghan
and Wo, Hongliang
and Gu, Yiqing
and Gu, Yimeng
and Ji, Li
and Wang, Wenbin
and Gou, Huiyang
and Shen, Yao
and Ying, Tianping
and Chen, Xiaolong
and Yang, Wenge
and Cao, Huibo
and Zheng, Changlin
and Zeng, Qiaoshi
and Guo, Jian-gang
and Zhao, Jun},
title={{Superconductivity in pressurized trilayer {La}$_{4}${Ni}$_{3}${O}$_{10-\delta}$ single crystals}},
journal={Nature},
year={2024},
month={Jul},
day={01},
volume={631},
number={8021},
pages={531-536},
issn={1476-4687},
doi={10.1038/s41586-024-07553-3},
url={https://doi.org/10.1038/s41586-024-07553-3}
}

@article{PhysRevB.109.144511,
  title = {{Theoretical analysis on the possibility of superconductivity in the trilayer Ruddlesden-Popper nickelate $\text{La}_{4}\text{Ni}_{3}\text{O}_{10}$ under pressure and its experimental examination: Comparison with $\text{La}_{3}\text{Ni}_{2}\text{O}_{7}$}},
  author = {Sakakibara, Hirofumi and Ochi, Masayuki and Nagata, Hibiki and Ueki, Yuta and Sakurai, Hiroya and Matsumoto, Ryo and Terashima, Kensei and Hirose, Keisuke and Ohta, Hiroto and Kato, Masaki and Takano, Yoshihiko and Kuroki, Kazuhiko},
  journal = {Phys. Rev. B},
  volume = {109},
  issue = {14},
  pages = {144511},
  numpages = {10},
  year = {2024},
  month = {Apr},
  publisher = {American Physical Society},
  doi = {10.1103/PhysRevB.109.144511},
  url = {https://link.aps.org/doi/10.1103/PhysRevB.109.144511}
}

@article{triqs_ctqmc_gull,
  title = {{Continuous-time quantum Monte Carlo algorithms for fermions}},
  author = {Emanuel Gull},
  year = {2008},
  doi = {10.3929/ethz-a-005722583},
  journal = {ETH}
}

@phdthesis{lewin_thesis,
  author = {Boehnke, Lewin Volker},
  title = {{Susceptibilities in materials with multiple strongly correlated orbitals}},
  school = {Universit\"{a}t Hamburg},
  year = {2015},
  url = {http://ediss.sub.uni-hamburg.de/volltexte/2015/7325/pdf/Dissertation.pdf}
}

@article{doi:10.7566/JPSJ.91.034002,
author = {Mizuno ,Ryota and Ochi ,Masayuki and Kuroki ,Kazuhiko},
title = {Simplification of the Local Full Vertex in the Impurity Problem in DMFT and Its Applications for the Nonlocal Correlation},
journal = {Journal of the Physical Society of Japan},
volume = {91},
number = {3},
pages = {034002},
year = {2022},
doi = {10.7566/JPSJ.91.034002},

URL = { 
    
        https://doi.org/10.7566/JPSJ.91.034002
    
    

},
eprint = { 
    
        https://doi.org/10.7566/JPSJ.91.034002
    
    

}
}

@article{PhysRevB.111.205136,
  title = {Improvement of the simplification method for the local two-particle full vertex towards precise frequency behavior},
  author = {Mizuno, Ryota and Kuroki, Kazuhiko and Ochi, Masayuki},
  journal = {Phys. Rev. B},
  volume = {111},
  issue = {20},
  pages = {205136},
  numpages = {13},
  year = {2025},
  month = {May},
  publisher = {American Physical Society},
  doi = {10.1103/PhysRevB.111.205136},
  url = {https://link.aps.org/doi/10.1103/PhysRevB.111.205136}
}


\end{document}